\documentclass[a4paper, 12pt]{article}
\usepackage{graphicx} 
\usepackage{amsmath}
\usepackage[top = 2cm, bottom = 2.2cm, left = 2.3cm, right = 2.3cm]{geometry}
\usepackage{multirow}
\usepackage[colorlinks,citecolor=blue,urlcolor=blue]{hyperref}
\usepackage[numbers]{natbib}
\usepackage{authblk}

\usepackage[utf8]{inputenc}
\usepackage[english]{babel}
\usepackage{float}
\usepackage{booktabs}
\usepackage{dsfont}

\usepackage{tikz}
\usetikzlibrary{positioning, shapes.geometric, fit, shapes.misc, arrows, shapes, quotes}

\title{Hidden Markov modelling of spatio-temporal dynamics of measles in 1750--1850 Finland}

\author[1]{Tiia-Maria Pasanen}
\author[1,2]{Jouni Helske}
\author[3,4]{Tarmo Ketola}
\affil[1]{Department of Mathematics and Statistics, University of Jyväskylä, tiia-maria.h.pasanen@jyu.fi}
\affil[2]{INVEST Research Flagship Centre, University of Turku}
\affil[3]{Department of Forestry, University of Helsinki}
\affil[4]{Department of Biological and Environmental Science, University of Jyväskylä}

\begin{document}
\maketitle

%
%



\begin{abstract}\noindent Real world spatio-temporal datasets, and phenomena related to them, are often challenging to visualise or gain a general overview of. In order to summarise information encompassed in such data, we combine two well known statistical modelling methods. To account for the spatial dimension, we use the intrinsic modification of the conditional autoregression, and incorporate it with the hidden Markov model, allowing the spatial patterns to vary over time. We apply our method to parish register data considering deaths caused by measles in Finland in 1750--1850, and gain novel insight of previously undiscovered infection dynamics. Five distinctive, reoccurring states, describing spatially and temporally differing infection burden and potential routes of spread, are identified. We also find that there is a change in the occurrences of the most typical spatial patterns circa 1812, possibly due to changes in communication networks after major administrative transformations in Finland.
\end{abstract}


\section{Introduction}\label{sec:intro}

Describing, visualising, and modelling spatio-temporal panel data is often challenging, particularly in high-dimensional settings with numerous time points and spatial areas. In this paper, we propose a Bayesian hidden Markov model with a state-dependent spatial correlation structure as an unsupervised learning method for decomposing the spatio-temporal variation in the data into relatively few interpretable latent components. We apply our approach by modelling the spatio-temporal occurrence of measles in Finland during the 18\textsuperscript{th} and 19\textsuperscript{th} centuries. We also explore how the large administrative changes---which Finnish society went through due to the annexation of Finland to Russia after the war between Sweden and Russia in 1809---altered the measles epidemics.

Wars, trade network changes, administrative border changes, and associated human movements are major upheavals in transmission networks of epidemics by affecting the connections between people. Human-to-human transmitted infections depend on the human networks, and closely connected populations are expected to indicate more similar infection patterns \citep{nitsch2025, ketola2021}. Therefore, new borders, on the one hand, and opened new connections, on the other, affect the spread of epidemics in the population. Vivid examples of such cases can be found in contemporary data showing, for example, the connection between the opening of trade and the spread of respiratory illnesses, and between increased mobility of people and the spread of HIV epidemics \citep{lin2022, oster2012}. Older data provide evidence, for example, of the effect of trade centres on historical plague epidemics \citep{yue2017}.

Throughout the history of mankind, armed conflicts and border changes between countries have been common, affecting the common welfare of people. For centuries, Finland had been part of Sweden, but in 1721, Russia annexed the southeastern parts of Finland, or Old Finland. After the Finnish War in 1808--1809, the rest of Finland became an autonomous grand duchy of Russia. In 1812, Old Finland was united with the new autonomous Finland, and the capital of Finland was moved from Turku on the southwestern coast to Helsinki on the southern coast, significantly affecting markets and trade connections \citep{ojala2017}. Being a remarkable transformation, the administrative change, however, left also many things unchanged. For example, the demographic records collected during the Swedish administration were still maintained in a similar way as before, due to which we have the data analysed here.

Finland has a long history in collecting demographic and health-related data, with local parishes systematically maintaining records of baptisms, burials, and causes of death from 1749 onwards. Even older data exist, but from this point on, the systematic data collection was mandatory, though not all collected data have survived to date \citep{pitkanen1977}. Although death diagnostics were symptom-based and determined by priests \citep[e.g.,][]{vuorinen1999}, the distinctive features of certain diseases allowed for relatively accurate diagnoses. One of such infections is measles, which we focus on. According to our data, measles accounted for approximately 2\% of all deaths during our study period from 1750 to 1850. The population of Finland was roughly $500{,}000$ in 1750, and grew to over 1.6 million by 1850 \citep{tilastollinen1899, voutilainen2020}. However, the healthcare system was primitive and virtually non-existent before the 19\textsuperscript{th} century, lagging behind in development compared to the other Nordic countries and Europe until the 20\textsuperscript{th} century \citep{saarivirta2009, saarivirta2012}.

Small population size coupled with hundreds of sparsely spaced small towns makes it difficult for epidemics to remain fully endemic \citep{keeling1997}. In such conditions of Finland, it is likely that fadeouts and reintroductions dominated the infection dynamics. This implies that changes in the transmission networks---routes and sources---could have strongly affected the infection dynamics, especially considering all the administrative changes affecting Finland during the era. These factors contributing to the sensitivity to transmission---small population, sparsely spaced towns, and low incidence---together with a large number of missing cases also make the Finnish data challenging to analyse.

Conditional autoregressive models (CAR) \citep{besag1974} and their variants, such as intrinsic conditional autoregressive models (ICAR), are frequently employed to describe areal data. Although it is possible to extend CAR models to spatio-temporal settings \citep{clayton1996}, these interaction models can become computationally intensive with high-dimensional data in which spatial and temporal effects are not separable \citep{knorr-held2000}. Approximate methods such as INLA \citep{rue2009} are computationally efficient alternatives to asymptotically exact Bayesian estimation via Markov chain Monte Carlo (MCMC), but these methods typically impose some restrictions on the model structure and the number of hyperparameters \citep{bakka2018}. The potential bias of these approximations is also challenging to quantify in specific cases, although in many typical settings the approximation bias has been found to be negligible \citep{ogden2016, rue2017}. However, visualising and interpreting the results of these interaction models may be complex: with time-varying spatial effects and a large number of time points, a substantial number of static maps are required to convey information about the estimated (or observed) spatial dynamics. This is at odds with our aim to communicate the main features and dominant patterns of the data in a compact and accessible form.

In life course research, hidden Markov models have been suggested as a probabilistic method to condense information of complex, multidomain life courses into a few interpretable latent states \citep{helske2018, helske2019}. In these applications, given a common data generating process, individuals are assumed to be independent, having their own latent trajectories. In contrast, in spatio-temporal panel data, individuals share a spatial dependency structure which should be accounted for.

Hidden Markov models have been used in epidemiological research also before. In \citet{amoros2020}, hidden Markov models are utilised in a spatio-temporal setting to detect influenza outbreaks, and in \citet{knorr-held2003}, they are used to identify meningococcal disease incidences. Both studies use two states to identify the endemic and epidemic, or endemic and hyperendemic, phases of the infections but do not aim to deeply analyse the underlying dynamics of the diseases. \citet{amoros2020} include the spatial dependence in the epidemic state as temporally independent ICAR components and omit it in the endemic state, whereas \citet{knorr-held2003} account for the spatial dependency as ICAR in both states but consider it as state and time-invariant. Employing more hidden states may aid in discovering latent phenomena related to the spatial and temporal patterns of epidemics. On the other hand, increasing the number of states increases computational demands, and eventually identifiability and interpretability issues, both in Bayesian and maximum likelihood settings. These discrete state hidden Markov models also fall outside the class of models supported by INLA framework.

By combining the hidden Markov model and the intrinsic conditional autoregression, we are able to acknowledge the spatial and temporal dependencies simultaneously and identify five easily interpretable disease states that altered in Finland in 1750--1850. We gain new insights into the Finnish disease and communication history and the behaviour of measles epidemics in an environment with a limited healthcare and sanitation system. Our study also demonstrates how sparse and scarce data with significant proportion of missing information can be analysed by combining well established methods in a novel way. The model can be applied to other diseases and datasets as well, possibly offering a new tool to learn from corresponding developing areas of modern world.\footnote{This text was originally published as a preprint (\url{https://arxiv.org/abs/2405.16885v1}).}\footnote{This text has been peer reviewed and published in Journal of Applied Statistics  (\url{https://doi.org/10.1080/02664763.2026.2634794}).}

\section{Data and methods}

In this section, first, we describe the data and their background in a historical and Finnish context. The rest of the chapter progresses from presenting the statistical components needed for our approach to presenting the main model itself. Subsection \ref{sec:hmm} introduces the hidden Markov models and subsection \ref{sec:icar} the intrinsic conditional autoregression. Subsection \ref{sec:mainmodel} represents in detail the main model and the necessary technical assumptions to estimate it, recalling the connection between the practice, given our application, and the statistical theory.

\subsection{Data}\label{sec:data}

We analyse deaths from measles recorded in the parish registers, using monthly data from January 1750 to December 1850 from $387$ different towns located in the southern half of Finland. The original records have count information on the deaths at the town level, but since the counts are generally low (99\% of all measles counts in our data are three or fewer) and there are no town-level population sizes available---only nationwide estimates \citep{voutilainen2020}---it is challenging to separate the actual differences between towns from those resulting from the varying population sizes.

To overcome these issues, we summarise the data into a more robust form. The observations we use are dichotomous, indicating whether at least one death caused by measles was recorded in each town. Thus, the data contains zeros for towns and months with no recorded deaths, and ones for those with recorded cases. Corresponding data, covering a different time span, has been used in \citet{pasanen2024}, which also describes the handling of the data in more detail. Other studies analysing data based on the same registers but covering different areas or time periods, or being aggregated to a different level, include, for example, \citet{ketola2021} and \citet{briga2021, briga2022}. \citet{pasanen2024} also implies that the dichotomisation of the observations results in conclusions corresponding to those of undichotomised data.

Although in general the cause of death records contain errors due to misdiagnoses, measles was likely diagnosed rather accurately because of its distinct visible symptoms. However, the dataset exhibits substantial missingness of records attributable to heterogeneity in record-keeping practices across towns \citep{pitkanen1977} and to the destruction of records in fires and other past events. This kind of missingness is unlikely to depend on the outcome variable itself.

Additionally, there are months during which no deaths for any cause have occurred due to the small population sizes---there were simply too few people for deaths to eventuate monthly in each town. In our data, such month-town observations are always coded as missing ones. In reality, this leads to overestimating the share of missing data. It is more likely that a missing observation between two months with detected cases means no deaths at all than missingness because of record keeping or data collection processes. Such missingness is dependent on the outcome variable and should be taken into consideration. Since the Finnish population is smaller the further back in time we look, the proportion of month-town combinations without any deaths in the older data is larger as well. Nevertheless, as the overall missingness is greater in the older records, it also becomes harder to identify and separate the general missingness of the records from the absence of deaths. Therefore, we assume that before the first observation for each town, the records are missing independently of the outcome (lack of records), while we account for potentially non-random, state-dependent missingness after the first observation in our proposed model.

\begin{figure}[h!]
    \centering
    \includegraphics{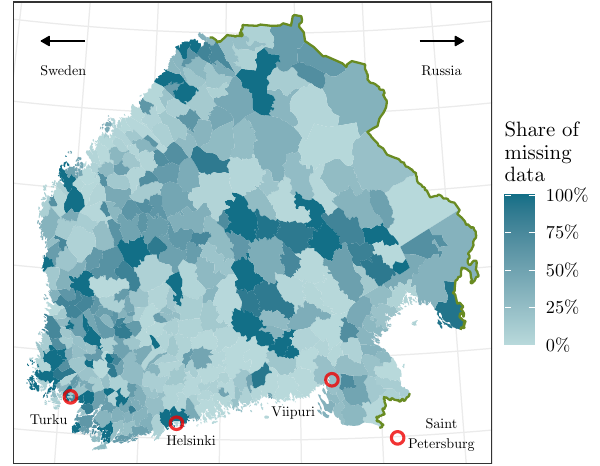}
    \caption{The percentage of all missing data across the years 1750--1850 in the towns on the study area, covering roughly the southern half of Finland. The red circles highlight the old capital Turku (until 1812), the new capital Helsinki (since 1812), the major trade city Viipuri which was united with the rest of Finland in 1812, and the then capital of Russia, Saint Petersburg. The borderline coloured with green denotes land border whereas the rest of the study area is bordered with water. The arrows denote the directions of the locations of Sweden and Russia.}
    \label{fig:missing}
\end{figure}

The study period covers $1212$ months, each having observations from $75$--$272$ towns per month, so no month has complete information from all towns. Only one town has complete records, whereas $44$ towns are missing all observations. The missingness is approximately 70\% at the beginning of the study period, decreasing steadily to about 20\% at the end of the time series, with an overall missingness rate approximately 40\%. However, when considering only data after the first observation in a town, the missingness rate is about 30\%. The spatial distribution of the overall missingness is illustrated in \autoref{fig:missing}.

In order to present our solution to analysing these data, next, we introduce the concepts of hidden Markov models and intrinsic conditional autoregression. The former compresses the time dependency into a relatively low number of nationwide discrete infection statuses, whereas the latter is responsible for considering the spatial connections.

\subsection{Hidden Markov Models}\label{sec:hmm}

Discrete time hidden Markov models (HMM) \citep{rabiner1989} are statistical models for temporal, or other sequentially ordered data, consisting of two essential components. There are, firstly, the hidden stochastic Markov process describing the unobserved, yet often interesting, part of the model, and secondly, the observed process dependent on the hidden component. With the word "state", often appearing in the terminology, we refer to some discrete, prevailing situations related to the phenomenon of interest. For example, there could be two states indicating the presence and absence of a disease, and the observed process would consist of the actual observations of the deaths caused by that disease. Next, we introduce the method in more detail.

The first component of the model is the hidden stochastic process $x_t$, $t=1,\dots,T,$ with the state space $\{1,\dots,S\}$, i.e., having values from $1$ to $S$. Usually the temporal dependency of the process $x_t$ encompasses only one previous time point so that it satisfies the first-order Markov property
\begin{equation}
    p(x_t | x_1, \dots, x_{t - 1}) = p(x_t | {x_{t - 1}}).
\end{equation}
The hidden process may also meet some other order Markov condition, meaning that the current state depends on as many preceding states as is the degree of the Markov chain, but we do not consider those cases here. For the states $s \in \{1,\dots,S\}$ we use consecutive numbering as labels  for simplicity. Between two preceding time points it is possible to stay at the current state or to switch to another state. The probabilities of different transitions between the states are collected into a transition matrix $A$, where each element $a_{s, s'} = P(x_t = s' | x_{t - 1} = s),$ where $s, s' \in \{1, \dots, S\}.$ Additionally, initial state probabilities $\rho_s = P(x_1 = s),\ s\in \{1, \dots, S\}$, describe the probabilities to start from each state at the first time point $t = 1$.

As a second component, we have the observed process $y_t$, $t = 1, \dots, T$, where $y_t$ depends only on the current state of the hidden process. In consequence, the observations are conditionally independent:
\begin{equation}\label{eq:hmm_y}
    p(y_t | y_1,\dots, y_{t - 1}, x_1, \dots, x_t) = p(y_t | x_t).
\end{equation}

Each state $x_t$ is associated with parameters $\theta_{x_t}$ of the observational distribution $p(y_t | x_t)$ and thus we can also write $p(y_t | x_t) = p(y_t | \theta_{x_t})$. Of the parameters $\theta_{x_t}$ some may depend on the value of the state $x_t$, while others may be shared between the states. Finally, we can define an emission matrix $\Omega$, where $\omega_{s, t} = p(y_t | \theta_{x_t = s})$. Now the HMM can be defined by the set $\{\Omega, A, \boldsymbol{\rho}\}$.

Often the hidden process and state-dependent parameters, describing the features of particular states, are the main interest of hidden Markov modelling, but since we do not directly observe the hidden process, we have to estimate it based on the observations gained from the non-hidden process. The general idea of the dependencies between the states, parameters and observations in HMMs is visualised in \autoref{fig:hmm}.

\begin{figure}[h!]
    \centering
    \tikzset{
    > = stealth,
    every node/.append style = {
        draw = none,
        text = black,
        minimum size = 1.2cm
    },
    every path/.append style = {
        arrows = ->,
        draw = black,
        fill = black
    },
    hidden/.style = {
        draw = lightgray,
        shape = circle,
        inner sep = 3pt
    },
    obs/.style = {
        draw = gray,
        shape = rectangle,
        inner sep = 4pt
    },
}
\tikz{
    \coordinate (input);
    \node[hidden] (x1) [right = of input] {$x_{1}$};
    \draw[olive] (input) to["\textcolor{olive}{$\rho$}" below = -.2cm] (x1);
    \node[obs] (y1) [above = of x1] {$y_{1}$};
    \path (x1) edge["\textcolor{magenta}{$\Omega$}" right = -.2cm, magenta] (y1);
    \node[xshift = -.5cm] (dots) [right = of x1] {\dots};
    \path (x1) edge["\textcolor{cyan}{$A$}" below = -.2cm, cyan] (dots);

    \node[hidden, xshift = -.5cm] (xt_0) [right = of dots] {$x_{t - 1}$};
    \node[obs] (yt_0) [above = of xt_0] {$y_{t - 1}$};
    \path (dots) edge["\textcolor{cyan}{$A$}" below = -.2cm, cyan] (xt_0);
    \path (xt_0) edge["\textcolor{magenta}{$\Omega$}" right = -.2cm, magenta] (yt_0);
    
    \node[hidden] (xt) [right = of xt_0] {$x_t$};
    \node[obs] (yt) [above = of xt] {$y_t$};
    \path (xt_0) edge["\textcolor{cyan}{$A$}" below = -.2cm, cyan] (xt);
    \path (xt) edge["\textcolor{magenta}{$\Omega$}" right = -.2cm, magenta] (yt);
    
    \node[hidden] (x2) [right = of xt] {$x_{t + 1}$};
    \node[obs] (y2) [above = of x2] {$y_{t + 1}$};
    \path (xt) edge["\textcolor{cyan}{$A$}" below = -.2cm, cyan] (x2);
    \path (x2) edge["\textcolor{magenta}{$\Omega$}" right = -.2cm, magenta] (y2);

    \node[xshift = -.5cm] (dots2) [right = of x2] {\dots};
    \path (x2) edge["\textcolor{cyan}{$A$}" below = -.2cm, cyan] (dots2);

    \node[hidden, xshift = -.5cm] (xT) [right = of dots2] {$x_{T}$};
    \path (dots2) edge["\textcolor{cyan}{$A$}" below = -.2cm, cyan] (xT);
    \node[obs] (yT) [above = of xT] {$y_{T}$};
    \path (xT) edge["\textcolor{magenta}{$\Omega$}" right = -.2cm, magenta] (yT);
}
    \caption{Directed graph of a hidden Markov model. The state $x_1$ in the first time point depends on the initial probabilities $\rho$, and the following transitions are determined by the transition probabilities in $A$. The observations $y_t$ depend on the current state $x_t$ via the emission probabilities in $\Omega$, and the current state $x_t$ in turn depends on the previous state $x_{t - 1}$.}
    \label{fig:hmm}
\end{figure}

While various methods have been suggested for estimating the number of states, $S$, for example, \citet{pohle2017}, typically it is treated as known, fixed value when analysing the HMM results. The interpretation ("labelling") of the hidden states is then based on the estimated model parameters $\{\Omega, A, \boldsymbol{\rho}\}$, and on the hidden state trajectory $x_1,\ldots,x_T$, given the observations $y_t$. As an example, consider HMM consisting of binary observations and two states, with $\boldsymbol{\rho} = (1, 0)^T$, $a_{1,1}=0.9$, $a_{2,2}=0.5$, $\omega_{1, t} = \textrm{Bernoulli(0.1)}$, and $\omega_{2, t} = \textrm{Bernoulli(0.8)}$. From these, we see that we always start from state one which is relatively persistent state ($a_{1,1}=0.9$) emitting mostly zeros, whereas we are more likely to observe ones when the hidden process is in state two. Based on this, if the observations were related, for example, to the occurrence of some disease, we could label the first state as endemic state and the second state as epidemic state. We could also gain further understanding from the estimated state trajectory, from which we could see, for example, that state one is more common during winter, whereas state two occurs only during summer months. This could lead us to conclude that the transmission probability of our disease is higher during summer than winter, and thus further guide us to find reasons for that.

To perform a fully Bayesian estimation of the HMM parameters and the corresponding latent states, we first run an MCMC targeting the marginal posterior of the model parameters, $p(\Omega, A, \rho | y)$, where $y = (y_1,\ldots,y_T)$. For this we need the marginal likelihood $p(y | \Omega, A, \rho)$, which can be computed with the forward part of the forward-backward algorithm \citep[and the references therein]{baum1966, rabiner1989}. The latent states $x = (x_1,\ldots, x_T)$ can then be sampled in the post-processing stage given the posterior samples of $\{\Omega, A, \rho\}$. This marginalisation approach commonly used with latent variable models allows much more efficient sampling than the approaches where the estimation is based on the joint posterior $p(x,\Omega, A, \rho)$, and is also strictly necessary when using efficient Hamiltonian Monte Carlo type of MCMC algorithms which rely on the differentiability of the posterior. 

The marginal likelihood approach is also convenient with respect to missing data under MAR. When computing the emission probabilities $\omega_{s, t}$, i.e., the conditional probabilities of the observations given the current state and the model parameters, we set $\omega_{s, t} = 1$ for all $s$ if the observation $y_t$ is missing. This means that as we do not gain any new information about the states due to the missing observation at time $t$, we essentially skip the contribution of $\omega_{s,t}$ at the marginal likelihood computation in forward algorithm at that time point \citep{zucchini2009}.

The likelihood and the posterior density of the hidden Markov models typically contain multiple modes. Some of these modes can be due to the label-switching problem \citep{jasra2005} but in complex models true local modes can also be present. In maximum likelihood setting the model parameters are often estimated by repeating the estimation several times with varying starting values for the optimisation algorithm, with the best solution declared as a global mode, while ignoring the other modes. In Bayesian setting multimodality of the posterior can pose both computational (convergence of the MCMC algorithms) and interpretability challenges. For example, it is difficult to interpret (label) and visualise the hidden states if the corresponding observational density is multimodal. These issues have to be taken into consideration when using HMMs.

The complexity of the response variables $y_t$ is not limited to scalars, but they can be multidimensional as well. In our case, they are vectors containing spatially dependent elements. This means that the states, as well as the parameters, should allow for spatial structure.

\subsection{Intrinsic conditional autoregressive models}\label{sec:icar}

Next, we introduce the intrinsic conditional autoregressive (ICAR) model, which is a special case of the conditional autoregressive (CAR) model \citep{besag1974}, especially meant for spatial analysis to depict the amount of dependency between different locations. These models are based on a spatial division of a larger area, employing a neighbourhood definition for the generated subareas, and combining both of these with a Gaussian distribution in order to quantify the amount of the spatial dependency. This method does not offer information on the direction of the effects, i.e., the causal relationships, but only on the total amount of dependency.

For the spatial division, denote each areal unit or site of a regional entity with $i = 1, \dots, N$, and correspondingly its neighbour with $j = 1, \dots, N$. To account for the contiguity of the sites, let $W$ be an $N \times N$ matrix for the binary neighbourhood so that the elements in it are
\begin{equation*}
    w_{i,j} = 
    \begin{cases}
        1, \text{ when sites $i$ and $j$ are neighbours, and $i \neq j$} \\
        0, \text{ otherwise.}
    \end{cases}
\end{equation*}
Additionally, denote with $D$ an $N \times N$ diagonal matrix indicating the amounts of neighbours of each site so that
\begin{equation*}
    d_{i,j} = 
    \begin{cases}
        \text{number of neighbours of site $i$,} \text{ when $i = j$} \\
        0, \text{ otherwise.}
    \end{cases}
\end{equation*}

Now, we employ a variable $\boldsymbol{\varphi} \in \mathds{R}^N$ to express the intensity of the dependency between each site and its neighbours. In case of the CAR model, we have
\begin{equation}\label{eq:car_distrib}
    \boldsymbol{\varphi} \sim \text{N}(0, Q^{-1}),
\end{equation}
where the precision matrix $Q$ has to be symmetric and positive definite for the distribution to have a proper joint probability density. This can be achieved with the matrices $W$ and $D$ by setting
\begin{equation}\label{eq:car_cov}
    Q = \tau D (I - \alpha D^{-1}W) = \tau (D - \alpha W),
\end{equation}
where $\tau \in \mathds{R}^+$ is a precision parameter of the variables $\varphi_i$, $\alpha \in (0,1)$ depicts the amount of spatial correlation, and $I$ is an identity matrix.

The normal distribution of formula \ref{eq:car_distrib} can also be written as
\begin{equation}\label{eq:car_dens}
    p(\boldsymbol{\varphi}) = (2 \pi )^{N/2} \left|(\tau(D - \alpha W))^{-1}\right|^{1/2} \text{exp}\left(-\frac{\tau}{2}\boldsymbol{\varphi}^\top(D - \alpha W) \boldsymbol{\varphi}\right)\!,
\end{equation}
where $\boldsymbol{\varphi}^\top$ denotes the transpose of $\boldsymbol{\varphi}$ and $|x|$ stands for the determinant of $x$. In case of ICAR model, we assume that $\alpha = 1$, which results in improper distribution, which cannot be used as a model for the data, but as a prior distribution for spatial dependency, as is done for example in the Besag York Mollié (BYM) models \citep{besag1991}. The parameters $\varphi_i$ are not identifiable without further constraints, as adding a constant to all of them does not change their density. A common way to avoid the identifiability problem is to restrict them to sum to zero over the sites.

Some reasons to employ the improper distribution of ICAR instead of the proper of the CAR, are the facts that, firstly, to account for a significant amount of spatial association, the parameter $\alpha$ most likely has to be close to one, and, secondly, the width of the posterior spatial pattern might be limited in the case of CAR \citep{banerjee2015}. Additionally, when employing the ICAR model, the specification $\alpha = 1$ obviates the need to calculate the determinant in \autoref{eq:car_dens} since it becomes constant, thus reducing the computation time even from days to hours \citep{morris2019}. Therefore, in some cases, it might be convenient to choose ICAR in the first place.

\subsection{Main model: HMM with ICAR}\label{sec:mainmodel}

In our approach, we combine the HMM and the spatial ICAR model, introduced above, thus enabling the HMM analysis of spatio-temporal data sets. We apply our method to the previously introduced historical Finnish parish register data to study the dynamics of the Finnish measles epidemics in 1750--1850. More precisely, we model the probability of observing at least one measles induced death in each town at each month.

As mentioned in \autoref{sec:data}, our data are missing a large number of observations. Under the assumption that they are missing at random, the HMM is a convenient approach to work with such imperfect data sets. Due to the hidden level, the observations do not depend on time, conditioned on the states, which is not the case when modelling dependencies only on the observed level. Now, the likelihood can be calculated without the missing observations by computing $p(y_t | x_t)$ over only the observed sites at time $t$ \citep{zucchini2009}. Thus, missing observations are naturally handled during model estimation, yet the model provides estimates for all sites and time points, with the missing cases influencing the posterior uncertainty of the estimates.

Using simulation experiments, \citep{yeh2012} showed that the bias due to not accounting non-ignorable missingness mechanism in HMMs can depend on the stability of the hidden states. Specifically, they found that the biases were smaller in cases of large self-transition probabilities (diagonal values of $A$). \citep{speekenbrink2021} studied HMMs in a case where missingness depends on the hidden state (and is thus missing not at random, MNAR). We follow their suggestion of jointly modelling $y_t$, the main outcome of interest, and binary missingness indicator $r_t$, assuming conditional independence of $y_t$ and $r_t$ given the hidden state $x_t$. Therefore, instead of simple MAR assumption, our model is also compatible with state-dependent MNAR.

We denote each site, i.e. town, with $i = 1, \dots, N$, and each time point, i.e. month, with $t = 1, \dots, T$. We model the probability of observing at least one death $y_{i,t}$ caused by measles in site $i$ and time point $t$, and the corresponding missingness indicator $r_{i,t}$, as 
\begin{align}\label{eq:model}
    y_{i,t} &\sim \text{Bernoulli}(\text{logit}^{-1}(\mu_{x_t} + \lambda_{i} + \varphi_{i,x_t} + \gamma_{t})),\\
    r_{i,t} &\sim \text{Bernoulli}(\text{logit}^{-1}(
    \xi_{x_t} + \beta_{x_t} t')),
\end{align}
where $t' = (t - 1) / (T - 1)$ and the latent, discrete states
\begin{equation}\label{eq:model_states}
    x_t | x_{t - 1} \sim \text{Categorical}(A_{x_{t - 1},.})
\end{equation}
represent the prevailing nationwide infection statuses. The notation $A_{x_{t - 1},.}$ means the row $x_{t - 1}$ including all columns of the transition matrix $A$. As we are using our method to study the occurrences and absences of deaths, Bernoulli distribution is a natural choice for the model. With other kinds of response variables also the distribution should be changed, for example Poisson distribution for counts. In formula \ref{eq:model}, which corresponds to the $p(y_t|x_t)$ in \autoref{eq:hmm_y}, the logit link function maps the combination of the explanatory terms onto a probability scale required by the Bernoulli distribution. Also the link function should be changed to an appropriate one when using a different distribution. The four explanatory terms in the model of deaths $y_{i,t}$ describe the state specific constants, the local constants, the intensity of spatial interaction between the site and its neighbours, and a monthly seasonal variation. The missingness probability is assumed to depend on the previous missingness $r_{i, t-1}$, current hidden state $x_t$, and the (scaled) calendar time $t'$. The formula \ref{eq:model_states} defines the connection between the underlying states $x_t$ and the transition matrix $A$. To simplify the notation, we occasionally use subscript $s$ to refer a case $x_t=s$, as in $\mu_s=\mu_{x_t = s}$, $s \in \{1,\dots,S\}$. 

The state specific constant $\mu_{x_t = s}$, $s \in \{1,\dots,S\}$, sets a base level for the nationwide probability of observing at least one death in each state $s$. In order to reduce the common, yet problematic, multimodality encountered with HMMs, we define the state specific constants so that they are monotonically increasing in a similar way as described in \citet{burkner2020}. Due to this, our states are ordered based on the general incidence level. In our case, to begin with, we define the constants for the first and last state, $\mu_1$ and $\mu_S$, respectively. We use these two to scale the ones for the other states. We do this by defining a simplex $\boldsymbol{m}$ of S components, i.e., a vector summing to one, with the first term $m_1$ fixed to zero, and defining the rest of the constants as
\begin{equation}
    \mu_s = \mu_1 + (\mu_S - \mu_1) \sum_{i = 1}^{s} m_i \text{ for all } s = 1, \dots, S.
\end{equation} 
As we are using a Bayesian approach, this allows us to employ informative priors for the constants of the first and last state, $\mu_1$ and $\mu_S$.

To allow local deviations from the nationwide base level, we add a site dependent term $\lambda_i$, which is constant over the states, i.e., over time. These local terms do not have a spatial dependency structure by definition, but they aim to capture any heterogeneous, time constant features of the sites, for example, differing population densities or communication intensities in our case. As we set the mean of the local constants to zero to improve model identifiability and MCMC sampling efficiency, we also gain a more straightforward interpretation of the state dependent constant $\mu_s$ as a nationwide average level.

The actual spatial dependence between the sites is introduced in the third term, $ \varphi_{i, x_t}$. The dependency is achieved via the ICAR structure outlined above, i.e., adapting the formulas \ref{eq:car_distrib} and \ref{eq:car_cov}, and letting $\boldsymbol{\varphi}_{s} \sim \text{N}(0, \sigma_{ \varphi, s}^2(D - W)^{-1})$, where $s \in \{1, \dots, S\}$. To employ the ICAR component, we need a definition for the neighbourhood. First of all, our sites are the towns based on the parishes, i.e., they are areas with borders. Here, two regions are considered as neighbours if they share a border. Other definitions, for example, based on the distance from the centers of the sites, could be used as well. In practice, the matrices $W$ and $D$ are constructed according to our neighbourhood definition in order to access the covariance of the spatial terms. For identifiability, we use sum-to-zero constraint on the state specific spatial terms $\varphi_{i, s}$, $s=1,\ldots,S$.
 
The final component $\gamma_t$ in the model is a monthly average deviation additional to all the other components, defined as
\begin{equation*}
    \gamma_{t} = \gamma_{m}, \quad m = 1 + (t - 1) \text{ mod } 12, \quad \sum_{m=1}^{12}\gamma_{m}=0.
\end{equation*}
This means that we assume that this seasonal term is non-stochastic in a sense that the monthly effects do not vary over years, and the sum of the monthly effects is zero \citep{durbin2012}.

To estimate the model we use a Bayesian approach, which on its behalf allows smooth handling of missing data, inclusion of prior information and interpretable uncertainty quantification. This procedure requires setting prior distributions for the unknown parameters, and for the full model they are

\begin{align*}
     \mu_1 &\sim \text{N}(-4.5, 0.25^2),\\
    \mu_S &\sim \text{N}(-1.75, 0.5^2)[\mu_1, ],\\
    \boldsymbol{m}_{2:S} &\sim \text{Dirichlet}(5_1, \dots, 5_{S - 1}), \\
    \boldsymbol{\varphi}_s &\sim \text{N}(0, \sigma_{ \varphi, s}^2 (D - W)^{-1}),  \text{ given } \frac{1}{N}\sum_{i=1}^N \varphi_{i,s} = 0,\\
    \lambda_{i} &\sim \text{N}(0, \sigma_{ \lambda}^2), \text{ given } \frac{1}{N}\sum_{i=1}^N \lambda_{i} = 0,  \\
    \sigma_\lambda &\sim \text{N}(0, 1)[0,],\\
    \sigma_\varphi &\sim \text{N}(0, 1)[0,],\\
    \gamma_{m} &\sim \text{N}(0, 1), \text{ given }  \frac{1}{12}\sum_{m=1}^{12} \gamma_{m} = 0, \\
    \boldsymbol{\rho} &\sim \text{Dirichlet}(1_1, \dots, 1_S), \text{ and}\\
    A_{s,.} &\sim \text{Dirichlet}(p_1, \dots, p_S), \text{ where } p_{s'} = \begin{cases}
        0.5 \text{, when } s' \neq s \\
        2S \text{, when }  s' = s,
    \end{cases} \\
\end{align*}
$s = 1, \dots, S$, $m = 1, \dots, 12$, $i = 1, \dots, N$, where $\text{N}(.,.)[z,]$ denotes a normal distribution truncated at $z$. In our case, the number of time points $T = 1212$ and the number of sites $N = 387$. For our final model, we set the number of states to be $S = 5$. 

These priors are mostly weakly informative and chosen to enhance the computational efficiency \citep{muth2018, banner2020}. In addition, these, together with the initial values for MCMC obtained from pilot runs aimed to locate the region of highest posterior density, are used to reduce the risk of the MCMC sampler wandering and getting trapped around minor modes of the complex posterior of HMMs. As a consequence of ignoring such regions of the posterior, the following results can underestimate the full posterior uncertainty.

\section{Results}

The model is estimated with MCMC using \textit{cmdstanr} \citep{gabry2022}, which is an R interface \citep{r2023} for the probabilistic programming language Stan for statistical inference \citep{stan2024}. Posterior samples are drawn using NUTS sampler \citep{hoffman2014, betancourt2018} with four chains. Each chain consists of $15{,}000$ iterations, the first $5{,}000$ being discarded as warm-up. The computation with parallel chains takes approximately 29 hours in total. The estimation is done on a supercomputer node with four cores of Xeon Gold 6230 $2.1$ GHz processors and $16$ GB of RAM. According to the MCMC diagnostics of the \textit{cmdstanr} \citep{vehtari2021}, the model converges without any divergences. The $\widehat{R}$ statistics are all below $1.004$, and the bulk and tail effective sample sizes are roughly between $1{,}100$ and $73{,}000$, with the smallest ones corresponding to the missingness parameters of $\xi$ and $\beta$ of State 2. The R and Stan codes and the data used for the analysis are available on GitHub (\url{https://github.com/tihepasa/spatialHMM}).

\begin{figure}[h!]
    \centering
    \includegraphics{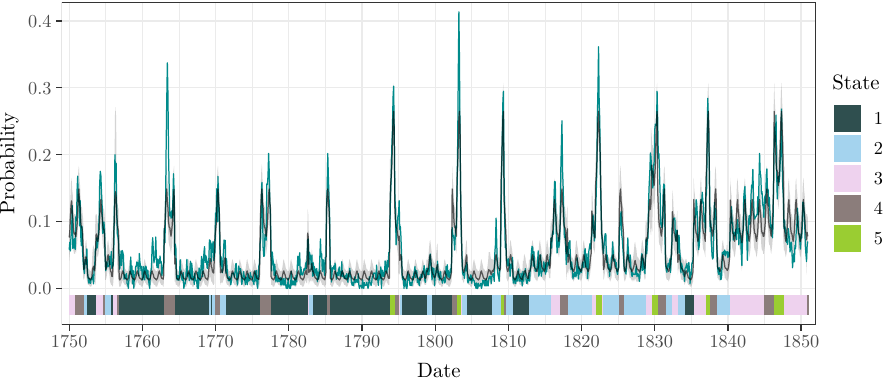}
    \caption{The turquoise line denotes the proportion of towns where at least one death caused by measles was observed in that month according to the data. The black line shows the expected value of the posterior predictive distribution of the nationwide proportion of towns with at least one death by measles. The grey area around the black line shows the 95\% posterior interval of the predicted average. The colours below the curve indicate the most likely hidden state at that time.}
    \label{fig:states_m}
\end{figure}

\autoref{fig:states_m} shows, for each time point $t$, the observed proportion of towns with at least one death caused by measles, and the posterior predictive mean and 95\% quantiles of this proportion based on our model. While the observed proportion is computed using only towns with non-missing observations at time $t$, the model-based estimates include all 387 towns. Specifically, at time $t$, for each posterior draw of the model parameters, we first draw a Bernoulli realisation $\tilde y_{i,t}$ with probability $ \text{logit}^{-1}(\mu_{s} + \lambda_{i} + \varphi_{i,s} + \gamma_{t})$ for all $i=1,\ldots,387$. We then compute the marginal mean of $\tilde y_{i,t}$ over $i$, and finally, the posterior predictive mean and quantiles are computed across all posterior draws.

It appears that State 1 is predominantly prevalent before the year 1813, while States 2, 3 and 5 are emphasised during the remainder of the study period. State 5 emerges for the first time in 1793, whereas States 2 and 3 have a few occurrences since the beginning. State 4, on the other hand, appears throughout the whole study period. In general, State 1 is the most likely during $45$\% of the months, State 2 during $22$\% of the months, and States 3 and 4 during $16$\% and $13$\% of the months, respectively, leaving State 5 prevailing during $5$\% of the months. The probabilities of the estimated most likely states to be the actual most likely states always range between $44-100$\%, with mean of $97$\%, less than $4$\% having probability smaller than $70$\%, and only three, randomly distributed in terms of time, falling below $50$\%.

\begin{figure}[h!]
    \centering
    \includegraphics{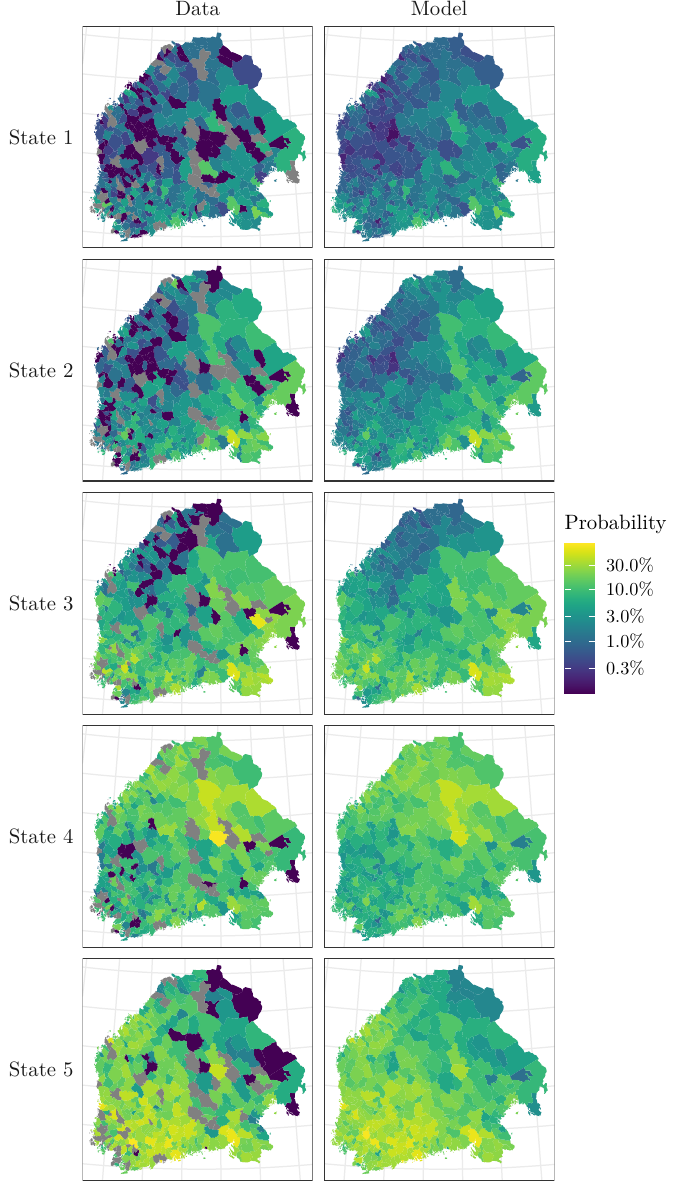}
    \caption{On the left column are the proportion of observations of at least one death caused by measles in each town as over the time points when each hidden state was most likely, and on the right column are the corresponding estimated probabilities logit$^{-1}(\mu_{s} + \lambda_i + \varphi_{i, s} + \gamma_t)$, also averaged over the the time points where state $s$ was most likely. The rows represent different states. Grey area indicates a site with only missing observations in months corresponding to that state. Note that the colour scale is logarithmic.}
    \label{fig:prob_m}
\end{figure}

The spatial dimension of the states can be seen in \autoref{fig:prob_m}, which presents the estimated local probabilities and the corresponding data of observing at least one death caused by measles in each state. The probabilities are computed in accordance with the model, as in the temporal case above, aggregated over time, and displayed on a logarithmic scale to facilitate easier comparison both between and within the states. The smallest probabilities are associated with State 1, which seems to be common during the periods when the proportion of the towns observing at least one death is low, see \autoref{fig:states_m}. In States 1, 2 and 3, the probabilities seem to be larger in the southeastern parts of the country, and as moving from State 1 to the others, the increased probabilities spread wider covering almost the whole study area in State 3. In State 4, the highest probabilities are located in the northern regions, and in State 5, they are concentrated in the southwestern half of the area.

The dark blue areas in the data column of \autoref{fig:prob_m} indicate that there are some towns where no deaths caused by measles were recorded, though it seems unlikely that any town would have been isolated enough to avoid the disease completely. Possible explanations for the lack of measles observations in these towns could be misdiagnoses or varying record-keeping practices \citep{pitkanen1977}.

\begin{figure}[!h]
    \centering
    \includegraphics{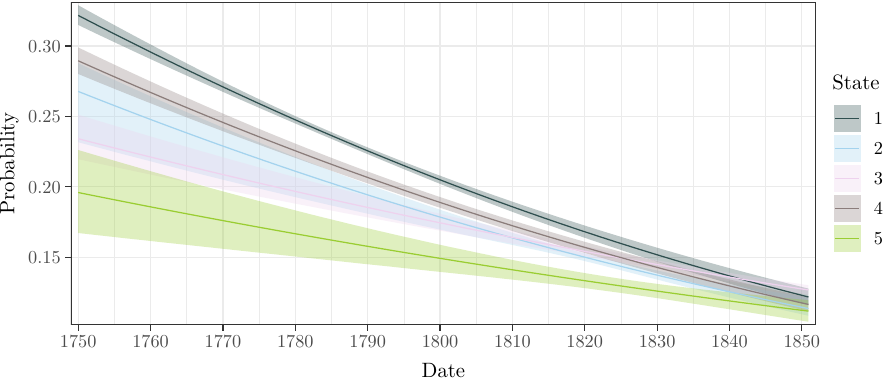}
    \caption{Estimated probability of an observation to be missing in each state over the time. The lines represent the posterior means and the shaded areas 95\% posterior intervals.}
    \label{fig:missingness_model}
\end{figure}

We estimated the probability of an observation to be missing in each state over the time, and the results are shown in \autoref{fig:missingness_model}. The missingness seems to be the most likely in State 5 and the least likely in State 1. In the beginning of our study period the differences between states are clear, whereas in the end the probabilities overlap. In concordance with the observed missingness, the estimated probability of missingness in each state also decreases with time.

Conditioning on the most likely hidden state at each time point, we also summarise the average death and missingness probabilities over towns and time points, and the corresponding observed proportions of death and missingness indicators $y$ and $r$, in \autoref{tab:data_vs_model}.  We see that on average our model performs well.

\begin{table}[h!]
\centering
\caption{\label{tab:data_vs_model} Observed proportions, posterior mean and 95\% quantiles of the death and missingness probabilities by most likely hidden state.}
\begin{tabular}{lrrrrrrrr}
\toprule
\multicolumn{1}{c}{ } & \multicolumn{4}{c}{Death} & \multicolumn{4}{c}{Missingness} \\
\cmidrule(l{3pt}r{3pt}){2-5} \cmidrule(l{3pt}r{3pt}){6-9}
State & Observed & Mean & 2.5\% & 97.5\% & Observed & Mean & 2.5\% & 97.5\%\\
\midrule
1 & 0.02 & 0.02 & 0.00 & 0.08 & 0.24 & 0.24 & 0.15 & 0.31\\
2 & 0.04 & 0.04 & 0.00 & 0.17 & 0.16 & 0.16 & 0.12 & 0.26\\
3 & 0.10 & 0.10 & 0.00 & 0.33 & 0.15 & 0.15 & 0.13 & 0.24\\
4 & 0.11 & 0.11 & 0.02 & 0.31 & 0.19 & 0.20 & 0.12 & 0.29\\
5 & 0.21 & 0.21 & 0.02 & 0.56 & 0.13 & 0.13 & 0.11 & 0.16\\
\bottomrule
\end{tabular}
\end{table}

According to the transition probabilities, shown in \autoref{tab:A}, it is most likely to remain in the current state. If we disregard this possibility, the most probable transitions are from State 1 to State 2, from State 2 to State 1, from State 3 to States 2 and 4, from 4 to State 2, and from State 5 to State 4. It should be remembered that these probabilities describe the overall transition probabilities instead of the likelinesses of the fast, abrupt monthly transitions. The estimates of the initial state probabilities $\rho_s$, along with other scalar parameters, can be found from \autoref{tab:statistics}. State 3 appears to be the starting point for the state trajectory in these data. However, the posterior distributions of the initial probabilities are quite wide. This is attributable to the fact that there is only one state trajectory to estimate, and an estimation based on a single observation does not provide much information.

\begin{table}[h!]
\centering
\caption{\label{tab:A}Posterior means and 95\% posterior intervals of the transition probabilities in the transition matrix $A$. Rows represent the states to transfer from and the columns the states to transfer to.}
\begin{tabular}{lccccc}
  \toprule
  & State 1 & State 2 & State 3 & State 4 & State 5\\
  \hline
  &&\\[-2.1ex]
  \multirow{2}{1.5cm}{State 1} & {0.97} & {0.01} & {0.01} & {0.01} & {0.00} \\
  & (0.95, 0.98) & (0.00, 0.03) & (0.00, 0.02) & (0.00, 0.02) & (0.00, 0.01) \\
  \multirow{2}{1.5cm}{State 2} & {0.04} & {0.92} & {0.02} & {0.01} & {0.00} \\
  & (0.02, 0.07) & (0.89, 0.95) & (0.01, 0.04) & (0.00, 0.03) & (0.00, 0.02) \\
  \multirow{2}{1.5cm}{State 3} & {0.01} & {0.03} & {0.91} & {0.03} & {0.02} \\
  & (0.00, 0.02) & (0.01, 0.06) & (0.87, 0.95) & (0.01, 0.06) & (0.01, 0.05) \\
  \multirow{2}{1.5cm}{State 4} & {0.03} & {0.05} & {0.02} & {0.89} & {0.02} \\
  & (0.01, 0.06) & (0.02, 0.09) & (0.00, 0.04) & (0.84, 0.94) & (0.00, 0.04) \\
  \multirow{2}{1.5cm}{State 5} & {0.01} & {0.01} & {0.04} & {0.07} & {0.87} \\
  & (0.00, 0.03) & (0.00, 0.04) & (0.01, 0.10) & (0.02, 0.14) & (0.79, 0.94) \\
   \bottomrule
\end{tabular}
\end{table}

\begin{table}[h!]
\centering
\caption{\label{tab:statistics}Posterior means and standard deviations, 95\% posterior intervals, bulk and tail effective sample sizes (ess) and $\widehat{R}$ statistics of the initial probabilities $\rho_s$, the state specific constants $\mu_s$, the deviation of local constants $\sigma_{\lambda}$, and the deviation parameter $\sigma_\varphi$ of the ICAR component.}
\begin{tabular}{rrrrrrrr}
  \toprule
 & mean & sd & 2.5\% & 97.5\% & ess bulk & ess tail & $\widehat{R}$ \\ 
  \midrule
$\rho_1$ & 0.17 & 0.14 & 0.00 & 0.52 & 47223 & 21092 & 1.00\\
$\rho_2$ & 0.17 & 0.14 & 0.00 & 0.53 & 28095 & 23330 & 1.00\\
$\rho_3$ & 0.33 & 0.18 & 0.05 & 0.72 & 32695 & 23143 & 1.00\\
$\rho_4$ & 0.17 & 0.14 & 0.01 & 0.52 & 52046 & 23485 & 1.00\\
$\rho_5$ & 0.17 & 0.14 & 0.01 & 0.52 & 54981 & 24539 & 1.00\\
$\mu_1$ & -4.63 & 0.05 & -4.73 & -4.53 & 8688 & 11541 & 1.00\\
$\mu_2$ & -3.93 & 0.05 & -4.02 & -3.84 & 9585 & 15357 & 1.00\\
$\mu_3$ & -2.74 & 0.03 & -2.80 & -2.67 & 20500 & 28782 & 1.00\\
$\mu_4$ & -2.39 & 0.03 & -2.45 & -2.34 & 28311 & 31093 & 1.00\\
$\mu_5$ & -1.66 & 0.03 & -1.72 & -1.59 & 21989 & 28406 & 1.00\\
$\sigma_{\phi_1}$ & 1.41 & 0.11 & 1.21 & 1.63 & 8132 & 15750 & 1.00\\
$\sigma_{\phi_2}$ & 1.02 & 0.09 & 0.84 & 1.21 & 3875 & 6601 & 1.00\\
$\sigma_{\phi_3}$ & 0.80 & 0.07 & 0.68 & 0.94 & 4994 & 9495 & 1.00\\
$\sigma_{\phi_4}$ & 0.59 & 0.07 & 0.47 & 0.73 & 3823 & 5619 & 1.00\\
$\sigma_{\phi_5}$ & 0.75 & 0.07 & 0.63 & 0.89 & 5432 & 11167 & 1.00\\
$\sigma_{\lambda}$ & 0.58 & 0.03 & 0.53 & 0.65 & 11795 & 20730 & 1.00\\
$\xi_1$ & -0.75 & 0.02 & -0.78 & -0.71 & 10165 & 18381 & 1.00\\
$\xi_2$ & -1.00 & 0.07 & -1.17 & -0.91 & 1858 & 1125 & 1.00\\
$\xi_3$ & -1.19 & 0.04 & -1.27 & -1.10 & 4733 & 2167 & 1.00\\
$\xi_4$ & -0.90 & 0.02 & -0.94 & -0.85 & 36124 & 28992& 1.00\\
$\xi_5$ & -1.42 & 0.10 & -1.61 & -1.23 & 20976 & 25885 & 1.00\\
$\beta_1$ & 1.23 & 0.04 & -1.32 & -1.15 & 11799 & 19076 & 1.00\\
$\beta_2$ & -1.05 & 0.09 & -1.19 & -0.83 & 1924 & 1148 & 1.00\\
$\beta_3$ & -0.74 & 0.05 & -0.85 & -0.65 & 4993 & 2215 & 1.00\\
$\beta_4$ & -1.13 & 0.04 & -1.21 & -1.05 & 36627 & 28818 & 1.00\\
$\beta_5$ & -0.66 & 0.13 & -0.90 & -0.41 & 22940 & 27358 & 1.00\\
   \bottomrule
\end{tabular}
\end{table}

The scalar constants $\mu_s$, controlling the base levels of the probabilities to observe at least one death caused by measles in each state, are all negative in logit-scale and differ between the states distinctively, see \autoref{tab:statistics}. When these are transformed into a probability scale, the base probabilities range from $1$\% in State 1 to $16$\% in State 5.

\begin{figure}[h!]
    \centering
    \includegraphics{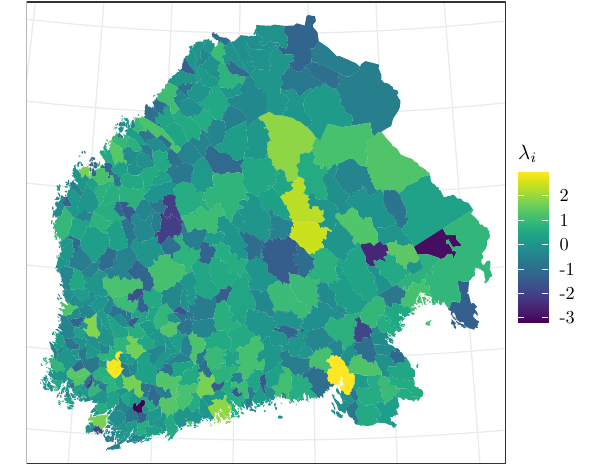}
    \caption{The posterior means of the local constants $\lambda_{i,s}$ by state.}
    \label{fig:lambda_i}
\end{figure}

The local constants $\lambda_i$, independent of the state, are shown in \autoref{fig:lambda_i}. As expected, there are no visible spatial patterns in these parameters, aimed to capture spatially independent differences of the sites. The deviation $\sigma_{\lambda}$ of the constants $\lambda_i$ is 0.58 with posterior interval [0.53, 0.65], see also \autoref{tab:statistics}, suggesting that there is some non-spatial heterogeneity between the towns. For instance, in State 5, the 5\% and 95\% quantiles for the posterior mean of the local probabilities $\text{logit}^{-1}(\mu_S + \lambda_i)$ are $(0.07,0.33)$ with median $0.17$.

\begin{figure}[hp]
    \centering
    \includegraphics{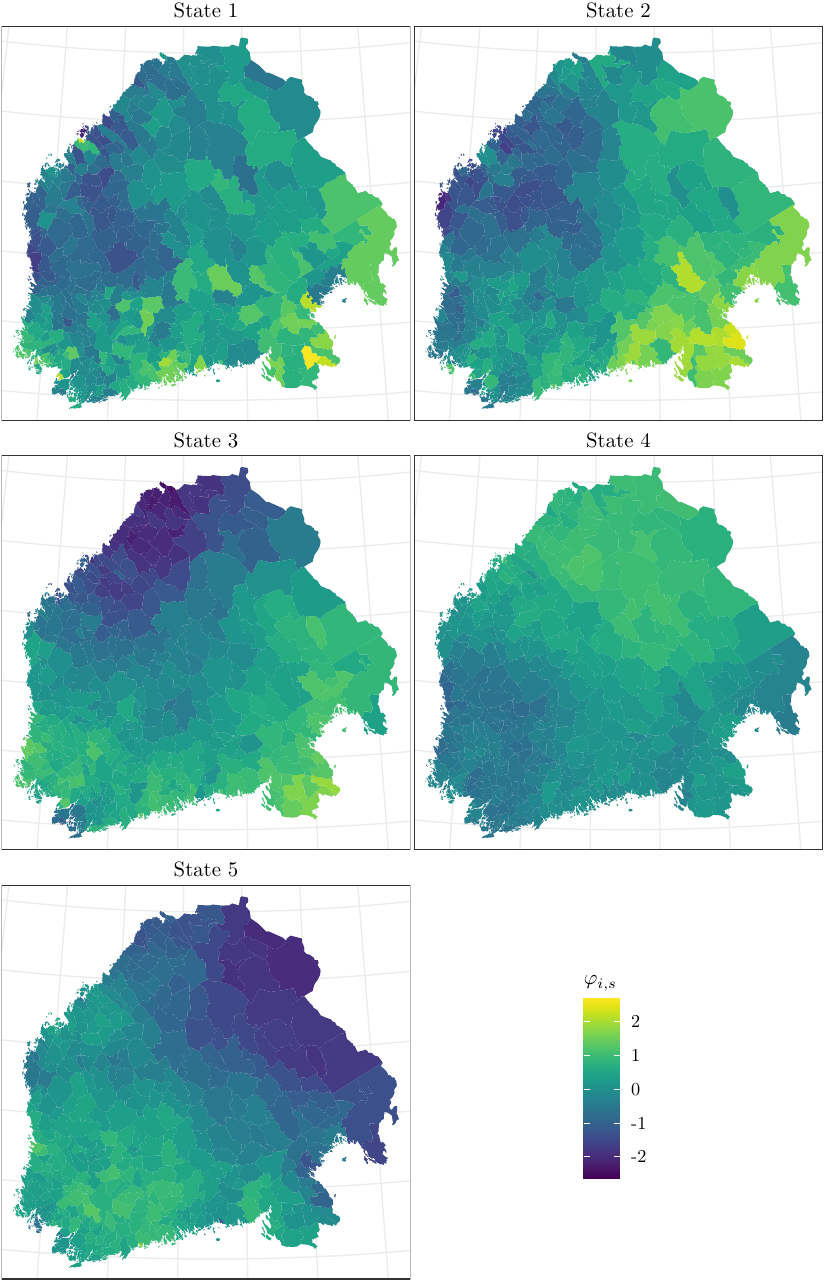}
    \caption{Posterior means of the spatial terms $\varphi_{i, s}$ by state.}
    \label{fig:phi_m}
\end{figure}

The spatial dependence in the spread of measles is accounted for by the spatial terms $\varphi_{i, s}$. The deviation parameter $\sigma_\varphi$ varies between states, from posterior mean of 1.4 [1.21, 1.63] in state 1 to 0.6 [0.47, 0.73] in state 4, see also \autoref{tab:statistics}. Combined with the information on the most likely states in each time point, we can deduce that the deviation of the spatial components $\varphi_{i, s}$ is greatest roughly before 1808. The state-specific spatial terms $\varphi_{i, s}$ are shown in \autoref{fig:phi_m}. The spatial patterns are similar to those of the estimated probabilities to observe at least one death caused by measles in \autoref{fig:prob_m}.

\begin{figure}[h!]
    \centering
    \includegraphics{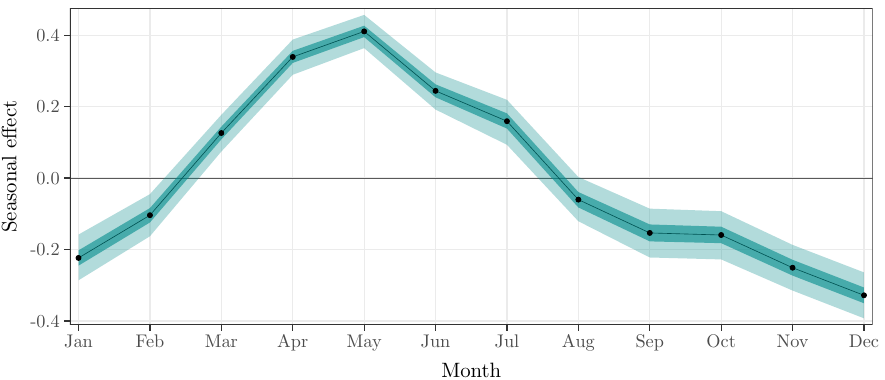}
    \caption{The black dots and line show the posterior mean of the monthly seasonal term $\gamma_t$. The dark and pale turquoise areas represent the $50$\% and $95$\% posterior intervals, respectively.}
    \label{fig:season}
\end{figure}

To capture the yearly temporal variation, the model includes a seasonal component $\gamma_t$, which is shown in \autoref{fig:season}. These monthly terms sum to zero over a year, i.e., they depict an average monthly variation additional to the other components of the model. The effect of the month increases from January to peak in May and decreases after that until December. The effect is positive, thereby increasing the probability of observing at least one death caused by measles, from March to July, and negative or decreasing the risk otherwise. A somewhat similar pattern, with a peak in May, was also identified in \citep{pasanen2024} for measles during 1820--1850 in Finland.

While our HMM was estimated as homogeneous in terms of the transition matrix $A$, \autoref{fig:states_m} suggests that the probability of observing deaths by measles changed around 1810. Before this, the hidden State 1 with the lowest baseline probabilities is the most prominent, after which the pattern of the hidden states appears to change considerably, both in terms of the baseline probability (transitioning from State 1 to State 2) and more frequent occurrences in the west coast (State 3).  Therefore, as an additional analysis, we aimed to identify the most probable point in time where the spatio-temporal dynamics of measles changed regimes. For this purpose, we estimated a two-state left-to-right HMM, where the observations consisted of a thousand posterior samples of the state trajectories of our main model. All trajectories were modelled jointly so that they shared the same model parameters (state-specific categorical emission probabilities and the single transition probability from State 1 to State 2), but each trajectory had a separate hidden state process. 

From the sampled state trajectories of this left-to-right model, we computed the distribution of the most probable change point, which was identified as November 1812 (with a 95\% posterior interval from October 1812 to February 1813). \autoref{tab:emission} displays the estimated emission matrix of this model. From this, we can conclude that the major changes were the increased probabilities of States 2, 3, and 5 and the decreased probability of State 1 after November 1812. This finding aligns well with the associated administrative changes that led to transferring the capital towards the east, and to altering the trade connections \citep{ojala2017}.

\begin{table}[ht]
\centering
\caption{\label{tab:emission}The emission matrix of the change point model including the posterior means and 95\% posterior intervals of the probabilities of being in each state before and after the change point.}
\vspace{0.2cm}
\begin{tabular}{lccccc}
  \toprule
  & State 1 & State 2 & State 3 & State 4 & State 5\\
  \hline
  &&\\[-2.1ex]
  \multirow{2}{1.5cm}{Before} 
  & {0.69} & {0.12} & {0.05} & {0.12} & {0.03} \\
  & (0.69, 0.69) & (0.12, 0.12) & (0.05, 0.05) & (0.12, 0.12) & (0.02, 0.03) \\ [2ex]
  \multirow{2}{1.5cm}{After} 
  & {0.04} & {0.39} & {0.34} & {0.14} & {0.09} \\
  & (0.04, 0.04) & (0.39, 0.39) & (0.34, 0.34) & (0.14, 0.14) & (0.09, 0.9) \\
   \bottomrule
\end{tabular}
\end{table}

Determining the number of states in HMM is not a trivial task, and there is no definitive procedure for that, even though some suggestions exist \citep[e.g.,][]{pohle2017}. In our case, we utilised five states, although a corresponding model was also estimated with four and six states. With six states there were several modes and convergence issues not only between but also within the chains. The model with four states, on the other hand, passed the convergence criteria but was not chosen here. Given that the states exhibited clear differences when there were five of them, we opted for the model with a greater number of hidden states. Presumably, this more detailed model also depicts the underlying phenomena more realistically in this context, since the epidemics are not discrete but continuous processes both in space and time.

Additionally, we compared the four-state and five-state versions of the main model, as well as a simpler model which has a state-independent $\sigma_\varphi$ and which does not have an explicit missingness model (see Section 2 in Supplementary material) using expected log predictive density (ELPD). This measures the goodness of the entire predictive distribution \citep{vehtari2017}. The commonly used approximate leave-one-out cross-validation resulted in high Pareto $\hat{k}$ values \citep{vehtari2017}, likely due to the multilevel structure with town-specific parameters and mixture nature of the observation probabilities. Therefore, we created ten replications of the original data, randomly leaving out additional 1\% (2838) observations in each. We then estimated candidate models and computed the pointwise log-likelihoods for the held-out data. \autoref{tab:elpd} shows the average and standard error of pairwise ELPD differences over the replicates when comparing different models. These results suggest a preference for the five-state models as well as for our main model. However, the average ELPD difference between the two five-state models is relatively small, which is to expected given their very similar performance for the death-occurrence outcome.

\begin{table}[ht]
\centering
\caption{\label{tab:elpd} Mean and standard error of pairwise ELPD differences over ten heldout samples for different models. Positive value means that the left-hand side model performed better. The simpler models are specified in Section 2 in Supplementary material.}
\begin{tabular}{l r r}
\hline
Comparison & ELPD$_{{\text{diff}}}$ & SE$_{\text{diff}}$ \\
\hline
Simpler S=5 vs. Simpler S=4 & 2.7 & 0.9 \\
Main S=5 vs. Main S=4 & 3.2 & 0.9 \\
Main S=5 vs. Simpler S=5 & 1.2 & 0.6 \\
\hline
\end{tabular}
\end{table}

For comparison, the supplementary material contains figures (Supplementary Figures 1--6) and tables (Supplementary Tables 1--2) for the four-state model, corresponding to the ones represented here for the five-state model in Figures \ref{fig:states_m}--\ref{fig:season} and Tables \ref{tab:A} and \ref{tab:statistics}. Based on Figures \ref{fig:states_m} and \ref{fig:prob_m} and Supplementary Figures 1 and 2, it seems that States 1 and 2 of the five-state model have merged into State 1 of the four-state model, but otherwise the states correspond to each other. Also in the four-state model, a change in the dynamics towards the end of the time series is visible (see Supplementary Figure 1). The figures and tables representing the results of the simpler models with four and five states are also included in Supplementary material.

\section{Discussion}

In this study, we demonstrated the usage of the hidden Markov models in spatio-temporal analysis. Our aim was to gain information on the historical Finnish measles epidemics as a spatial and temporal process by modelling the probability of observing at least one death caused by measles in each town and month. The proposed HMM approach allowed us to summarise the information of data that are otherwise complicated and challenging to understand.

Using reported fatalities of measles in Finland throughout 1750--1850, and the Bayesian hidden Markov model with state-dependent spatial correlation structure, we identified five reoccurring epidemic states in Finland. The epidemics were characterised by two states of low burden of infection (States 1 and 2), and three states of higher infectious burden (States 3, 4 and 5), covering different parts of Finland (see \autoref{fig:prob_m}).

Our analyses revealed also a distinctive change in the epidemic dynamics between November 1812 and February 1813, matching well with the changes caused by the war between Sweden and Russia over the area of Finland in 1808--1809, and the imposed large administrative changes, including the transfer of the capital from Turku to Helsinki, in 1812. These transformations strengthened significantly the eastern influence and trade connections \citep{ojala2017}. The most notable change in the estimated state trajectory after the annexation of Finland by Russia was the replacement of State 1 with a generally higher death observation probability of State 2 as a non-epidemic state. Also the likelihood of States 3 and 5 increased, reflecting the higher infection burden at southern Finland and Karelian Isthmus, or south-western Finland, respectively. The view of the strengthened eastern influence is further supported by the increase of the infection burden especially in the eastern towns during the low epidemic states (\autoref{fig:prob_m}), possibly indicating persistent infectious pressure from the nearby, then capital of Russia, Saint Petersburg (see \autoref{fig:missing}). Overall, infections became more common across whole Finland after the war.

The dynamics of measles and other contagious infections are sensitive to population size \citep{keeling1997}. Due to the clear abrupt change in infection dynamics, we can perhaps safely rule out the role of the gradual population increase \citep{voutilainen2020} here as the main driver of the change of the epidemics. Further examination considering particularly the spatial effect of the population sizes remains as a future work due to the missing local population size information. Especially sparsely populated areas without endemic measles infections follow the changes of adjacent larger populations \citep{keeling1997, grenfell1997, rohani2003, ketola2021}. Epidemics are affected by transport networks, geographical obstacles and borders \citep{vora2008}. Finland and Sweden are separated by a large geographical barrier: The Gulf of Bothnia is 80--150 kilometres wide and 700 kilometres long, and obviously restricting the movement of people---and infections. Sea on the other side and the Russian border on the other side seem to have kept the infectious burden smaller in pre-war Finland compared with the post-war era when contacts to Russia and elsewhere increased strongly \citep{ojala2017}. During the history of nations, the borders have changed due to wars and the directions of transport and trade have shifted frequently, but the dynamics of epidemics due to such regime shifts have not been closely followed. Luckily, in our case, the administrative changes did not modify the data collection procedures of parishes so that we had data of the deaths available. This allowed us to study more closely the dynamics of epidemics due to changing transmission networks.

When it comes to the local transmission routes inside Finland, the definition of the proximity between the towns plays a crucial role. Other relevant definitions for the neighbourhood could be road or water connections instead of or in addition to the border sharing we used. As during longer time periods the areal structure is rarely static due to wars and other environmental changes, it would be interesting to extend the model into a case in which the spatial structure and neighbours can vary in time. This would require some further specifications for the state-specific spatial coefficients, as allowing several spatial structures within one state could result in having not only $S$, but the number of different structures times $S$, states. To set the specifications and try this kind of approach in practice, one should have further information about the spatial structure and the changes in it, but technically it could be implemented with our model. Additionally, within a fixed spatial structure one could estimate if some neighbouring towns did not interact at all so that there were actually discontinuities, as in \citet{balocchi2019}.

As mentioned in \autoref{sec:intro}, there are several ways to model and estimate spatio-temporal connections. It would be natural to consider the phenomena as a continuous process instead of merely inspecting a collection of discrete states. One attractive way for this kind of more detailed analysis would be to fit a model without any hidden states, using random walks so that $y_{i,t} \sim \text{Bernoulli}(\mu_t + \lambda_i + \varphi_{i,t} + \gamma_t)$, where $\mu_t \sim \text{N}(\mu_{t - 1}, \sigma_\mu^2)$, and $\boldsymbol{\varphi}_{t} \sim \text{N}(\boldsymbol{\varphi}_{t - 1}, \sigma_\varphi(D - W)^{-1})$. In this case the spatio-temporal parameter $\varphi_{i,t}$ may be seen as an interaction parameter of the temporal and spatial terms, as described for example in \citet{clayton1996}. Given the large number of time points, spatial units, and iterations needed for the convergence of the MCMC, this kind of model is computationally infeasible to estimate in our case using MCMC, although approximate solutions using INLA would have likely been obtainable. However, the downside of such models is that due to the parameter profusion, the results would not be as interpretable or offer such a straightforward insight as our hidden Markov model.

We allowed for simplified temporal dependency structure for the model via the parameters that depend on the state. Our method is especially convenient when there is a reason to suspect a presence of recurring patterns in the underlying phenomena. In other words, as we allow all potential transitions between our states so that it is possible to return to any of the previously occurred states later in time, also the latent process consists of repeated patterns. This enables finding the spatio-temporal epidemic dynamics and transmission routes illustrated by the repeated peak and fade-out patterns. In addition to the state-specific constants $\mu_s$ and spatial terms $\varphi_{i,t}$ and the corresponding deviation parameters $\sigma_\varphi$, also the local constant $\lambda_i$, their standard deviations, and the seasonal term $\gamma_t$ could depend on state if seen appropriate. Our attempts to include the state dependent local constants and seasonal terms led to some identifiability issues. The same was also true for an alternative model in which, instead of the state-dependent $\mu_s$, the model included a temporal trend $\mu_t$ which evolved continuously over time (in which case the states describe only changes in the spatial patterns). Thus, the inclusion of these further complexities in our model remains as a future work.

Our data have a large proportion of missing observations due to different reasons, for example, register keeping practices and data preservation processes. The small population sizes cause additional missingness because for the towns the months without any registered deaths are coded as missing even though it is possible that in reality no one died, but we have no means to know that with the information we have currently available. As a result, we aggregated the data and modelled the missingness to enhance the reliability of our analysis. In this way, we were also able to gain information on the probability of missingness in different states during the hundred-year study period.

While our focus was on binary data, other observational distributions could be used and mixed, for example, a joint model of Poisson and Gaussian responses. There are previous indications of the profitability of the dichotomisation considering the last 30 years of our data \citep{pasanen2024}. The model may also be extended to a case with exogenous covariates for the observational distributions or for the transition probabilities.

\section*{Acknowledgements}
This research was supported by the Research Council of Finland (grant numbers 331817, 355153, 345546, and 278751), Emil Aaltonen Foundation, the Finnish Cultural Foundation, and INVEST Research Flagship Centre. 

The authors wish to acknowledge CSC – IT Center for Science, Finland, for computational resources. The authors also thank Virpi Lummaa and Harri Högmander.

\bibliography{bibliography}

@article{amoros2020,
  title={A spatio-temporal hierarchical {M}arkov switching model for the early detection of influenza outbreaks},
  author={Amor{\'o}s, Rub{\'e}n and Conesa, David and L{\'o}pez-Qu{\'\i}lez, Antonio and Martinez-Beneito, Miguel-Angel},
  journal={Stochastic environmental research and risk assessment},
  volume={34},
  number={2},
  pages={275--292},
  year={2020},
  publisher={Springer},
  doi = {https://doi.org/10.1007/s00477-020-01773-5}
}

@article{bakka2018,
author = {Bakka, Haakon and Rue, Håvard and Fuglstad, Geir-Arne and Riebler, Andrea and Bolin, David and Illian, Janine and Krainski, Elias and Simpson, Daniel and Lindgren, Finn},
title = {{Spatial modeling with R-INLA: A review}},
journal = {WIREs Computational Statistics},
volume = {10},
number = {6},
pages = {e1443},
doi = {https://doi.org/10.1002/wics.1443},
year = {2018}
}

@article{balocchi2019,
  title={{Spatial modeling of trends in crime over time in Philadelphia}},
  author={Balocchi, Cecilia and Jensen, Shane T},
  journal={The Annals of Applied Statistics},
  volume={13},
  number={4},
  pages={2235--2259},
  year={2019},
  publisher={JSTOR},
  doi = {https://doi.org/10.1214/19-aoas1280}
}

@book{banerjee2015,
  title={{Hierarchical Modeling and Analysis for Spatial Data, Second Edition}},
  author={Banerjee, S. and Carlin, B.P. and Gelfand, A.E.},
  isbn={9781439819173},
  lccn={2014451631},
  series={Chapman \& Hall/CRC Monographs on Statistics \& Applied Probability},
  year={2015},
  publisher={Taylor \& Francis},
  doi = {https://doi.org/10.1201/b17115-10}
}

@article{banner2020,
author = {Banner, Katharine M. and Irvine, Kathryn M. and Rodhouse, Thomas J.},
title = {{The use of Bayesian priors in Ecology: The good, the bad and the not great}},
journal = {Methods in Ecology and Evolution},
volume = {11},
number = {8},
pages = {882-889},
doi = {https://doi.org/10.1111/2041-210X.13407},
year = {2020}
}

@article{baum1966,
  title={Statistical inference for probabilistic functions of finite state {M}arkov chains},
  author={Baum, Leonard E and Petrie, Ted},
  journal={The Annals of Mathematical Statistics},
  volume={37},
  number={6},
  pages={1554--1563},
  year={1966},
  publisher={JSTOR},
  doi = {https://doi.org/10.1214/aoms/1177699147}
}

@article{besag1974,
author = {Besag, Julian},
title = {Spatial Interaction and the Statistical Analysis of Lattice Systems},
journal = {Journal of the Royal Statistical Society: Series B (Methodological)},
volume = {36},
number = {2},
pages = {192-225},
doi = {https://doi.org/10.1111/j.2517-6161.1974.tb00999.x},
year = {1974}
}

@article{besag1991,
  title={Bayesian image restoration, with two applications in spatial statistics},
  author={Besag, Julian and York, Jeremy and Molli{\'e}, Annie},
  journal={Annals of the Institute of Statistical Mathematics},
  volume={43},
  pages={1--20},
  year={1991},
  publisher={Springer},
  doi = {https://doi.org/10.1007/BF00116466}
}

@article{betancourt2018,
      title={A Conceptual Introduction to {H}amiltonian {M}onte {C}arlo}, 
      author={Michael Betancourt},
      year={2018},
      journal={arXiv},
      primaryClass={stat.ME},
      note={[Preprint]. Posted July 16, 2018 [accessed May 3, 2024]},
      doi = {https://doi.org/10.48550/arXiv.1701.02434}
}

@article{briga2021,
  title={The seasonality of three childhood infections in a pre-industrial society without schools},
  author={Briga, Michael and Ukonaho, Susanna and Pettay, Jenni E and Taylor, Robert J and Ketola, Tarmo and Lummaa, Virpi},
  journal={medRxiv},
  year={2021},
  publisher={Cold Spring Harbor Laboratory Press},
  note = {[Preprint]. Posted October 18, 2021 [accessed September 29, 2023]},
  doi = {https://doi.org/10.1101/2021.10.08.21264734}
}

@article{briga2022,
  title = {The epidemic dynamics of three childhood infections and the impact of first vaccination in 18th and 19th century {F}inland},
  author = {Briga, Michael and Ketola, Tarmo and Lummaa, Virpi},
  journal = {medRxiv},
  note = {[Preprint]. Posted October 31, 2022 [accessed May 25, 2023]},
  doi = {https://doi.org/10.1101/2022.10.30.22281707},
  year = {2022}
}

@article{burkner2020,
author = {Bürkner, Paul-Christian and Charpentier, Emmanuel},
title = {{Modelling monotonic effects of ordinal predictors in Bayesian regression models}},
journal = {British Journal of Mathematical and Statistical Psychology},
volume = {73},
number = {3},
pages = {420-451},
doi = {https://doi.org/10.1111/bmsp.12195},
year = {2020}
}

@incollection{clayton1996,
    author = {Clayton, D. and Bernardinelli, L.},
    isbn = {9780192622358},
    title = "{Bayesian methods for mapping disease risk}",
    booktitle = "{Geographical and Environmental Epidemiology: Methods for Small Area Studies}",
    publisher = {Oxford University Press},
    year = {1996},
    month = {07},
    doi = {https://doi.org/10.1093/acprof:oso/9780192622358.003.0018}
}

@book{durbin2012,
  title={Time series analysis by state space methods},
  author={Durbin, James and Koopman, Siem Jan},
  year={2012},
  publisher={Oxford university press},
  doi = {https://doi.org/10.1093/acprof:oso/9780199641178.001.0001}
}

@Manual{gabry2022,
    title = {cmdstanr: R Interface to 'CmdStan'},
    author = {Jonah Gabry and Rok Češnovar},
    year = {2022},
    note = {\url{https://mc-stan.org/cmdstanr/}, \url{https://discourse.mc-stan.org}},
  }

@article{grenfell1997,
  title={{(Meta)population dynamics of infectious diseases}},
  author={Grenfell, Bryan and Harwood, John},
  journal={Trends in Ecology \& Evolution},
  volume={12},
  number={10},
  pages={395--399},
  year={1997},
  publisher={Elsevier},
  doi = {https://doi.org/10.1016/S0169-5347(97)01174-9}
}

@incollection{helske2018,
  title={Combining sequence analysis and hidden {M}arkov models in the analysis of complex life sequence data},
  author={Helske, Satu and Helske, Jouni and Eerola, Mervi},
  booktitle={Sequence Analysis and Related Approaches. Life Course Research and Social Policies},
  editor = {Ritschard, G. and Studer, M.},
  doi = {https://doi.org/10.1007/978-3-319-95420-2_11},
  pages={185--200},
  year={2018},
  volume = {10},
  publisher={Springer International Publishing}
}

@article{helske2019,
 title={Mixture Hidden {M}arkov Models for Sequence Data: The {seqHMM} Package in {R}},
 volume={88},
 doi={https://doi.org/10.18637/jss.v088.i03},
 number={3},
 journal={Journal of Statistical Software},
 author={Helske, Satu and Helske, Jouni},
 year={2019},
 pages={1–32}
}

@article{hoffman2014,
  title={The {No-U-Turn} Sampler: Adaptively Setting Path Lengths in {H}amiltonian {M}onte {C}arlo.},
  author={Hoffman, Matthew D and Gelman, Andrew},
  journal={Journal of Machine Learning Research},
  volume={15},
  number={1},
  pages={1593--1623},
  year={2014}
}

@article{jasra2005,
author = {A. Jasra and C. C. Holmes and D. A. Stephens},
title = {Markov Chain {M}onte {C}arlo Methods and the Label Switching Problem in {B}ayesian Mixture Modeling},
volume = {20},
journal = {Statistical Science},
number = {1},
publisher = {Institute of Mathematical Statistics},
pages = {50 -- 67},
year = {2005},
doi = {https://doi.org/10.1214/088342305000000016}
}

@article{keeling1997,
author = {M. J. Keeling  and B. T. Grenfell },
title = {Disease Extinction and Community Size: Modeling the Persistence of Measles},
journal = {Science},
volume = {275},
number = {5296},
pages = {65-67},
year = {1997},
doi = {https://doi.org/10.1126/science.275.5296.65}
}

@article{ketola2021,
    author = {Ketola, Tarmo and Briga, Michael and Honkola, Terhi and Lummaa, Virpi},
    title = {{Town population size and structuring into villages and households drive infectious disease risks in pre-healthcare Finland}},
    journal = {Proceedings of the Royal Society B: Biological Sciences},
    volume = {288},
    number = {1949},
    pages = {20210356},
    year = {2021},
    doi = {https://doi.org/10.1098/rspb.2021.0356}
}

@article{knorr-held2000,
author = {Knorr-Held, Leonhard},
title = {Bayesian modelling of inseparable space-time variation in disease risk},
journal = {Statistics in Medicine},
volume = {19},
number = {17-18},
pages = {2555-2567},
year = {2000},
url = {https://doi.org/10.1002/1097-0258(20000915/30)19:17/18<2555::AID-SIM587>3.0.CO;2-%23}
}

@article{knorr-held2003,
  title={A hierarchical model for space--time surveillance data on meningococcal disease incidence},
  author={Knorr-Held, Leonhard and Richardson, Sylvia},
  journal={Journal of the Royal Statistical Society Series C: Applied Statistics},
  volume={52},
  number={2},
  pages={169--183},
  year={2003},
  publisher={Oxford University Press},
  doi = {https://doi.org/10.1111/1467-9876.00396}
}

@article{lin2022,
  title= {{How trade affects pandemics? Evidence from severe acute respiratory syndromes in 2003}},
  author={Lin, Faqin and Wang, Xiaosong and Zhou, Mohan},
  journal={The World Economy},
  volume={45},
  number={7},
  pages={2270--2283},
  year={2022},
  publisher={Wiley Online Library},
  doi = {https://doi.org/10.1111/twec.13127}
}

@article{morris2019,
title = {{Bayesian hierarchical spatial models: Implementing the Besag York Mollié model in stan}},
journal = {Spatial and Spatio-temporal Epidemiology},
volume = {31},
pages = {100301},
year = {2019},
issn = {1877-5845},
doi = {https://doi.org/10.1016/j.sste.2019.100301},
author = {Mitzi Morris and Katherine Wheeler-Martin and Dan Simpson and Stephen J. Mooney and Andrew Gelman and Charles DiMaggio},
}

@article{muth2018,
  title={User-friendly {B}ayesian regression modeling: A tutorial with rstanarm and shinystan},
  author={Muth, Chelsea and Oravecz, Zita and Gabry, Jonah},
  journal={The Quantitative Methods for Psychology},
  volume={14},
  number={2},
  pages={99--119},
  year={2018},
  publisher={The Quantitative Methods for Psychology},
  doi = {https://doi.org/10.20982/tqmp.14.2.p099}
}

@article{nitsch2025,
  title= {{The spatial distribution of pertussis, but not measles or smallpox, in pre-industrial Finland matches dialects}},
  author={Nitsch, A{\"\i}da and Lummaa, Virpi and Ketola, Tarmo and Honkola, Terhi and Vesakoski, Outi and Briga, Michael},
  journal={iScience},
  volume={28},
  number={6},
  year={2025},
  publisher={Elsevier},
  doi = {https://doi.org/10.1016/j.isci.2025.112530}
}

@article{ogden2016,
      title={On asymptotic validity of naive inference with an approximate likelihood}, 
      author={Helen Ogden},
      journal={Biometrika},
      volume={104},
      number={1},
      pages={153--164},
      year={2017},
      publisher={Oxford University Press},
      doi = {https://doi.org/10.1093/biomet/asx002}
}

@article{ojala2017,
  title={{Navigation Acts and the integration of North Baltic shipping in the early nineteenth century}},
  author={Ojala, Jari and R{\"a}ih{\"a}, Antti},
  journal={International Journal of Maritime History},
  volume={29},
  number={1},
  pages={26--43},
  year={2017},
  publisher={Sage Publications Sage UK: London, England},
  doi = {https://doi.org/10.1177/0843871416678166}
}

@article{oster2012,
  title={{Routes of infection: exports and HIV incidence in Sub-Saharan Africa}},
  author={Oster, Emily},
  journal={Journal of the European Economic Association},
  volume={10},
  number={5},
  pages={1025--1058},
  year={2012},
  publisher={Oxford University Press},
  doi = {https://doi.org/10.1111/j.1542-4774.2012.01075.x}
}

@article{pasanen2024,
  title={Spatio-temporal modeling of co-dynamics of smallpox, measles, and pertussis in pre-healthcare {F}inland},
  author={Pasanen, Tiia-Maria and Helske, Jouni and H{\"o}gmander, Harri and Ketola, Tarmo},
  journal={PeerJ},
  volume={12},
  pages={e18155},
  year={2024},
  publisher={PeerJ Inc.},
  doi = {https://doi.org/10.7717/peerj.18155}
}

@article{pitkanen1977,
author = {Kari Pitk{\"a}nen},
title = {{The reliability of the registration of births and deaths in Finland in the eighteenth and nineteenth centuries: Some examples}},
journal = {Scandinavian Economic History Review},
volume = {25},
number = {2},
pages = {138-159},
year  = {1977},
publisher = {Routledge},
doi = {https://doi.org/10.1080/03585522.1977.10407878}
}

@article{pohle2017,
  title={Selecting the number of states in hidden {M}arkov models: pragmatic solutions illustrated using animal movement},
  author={Pohle, Jennifer and Langrock, Roland and Van Beest, Floris M and Schmidt, Niels Martin},
  journal={Journal of Agricultural, Biological and Environmental Statistics},
  volume={22},
  pages={270--293},
  year={2017},
  publisher={Springer},
  doi = {https://doi.org/10.1007/s13253-017-0283-8}
}

@Manual{r2023,
    title = {R: A Language and Environment for Statistical Computing},
    author = {{R Core Team}},
    organization = {R Foundation for Statistical Computing},
    address = {Vienna, Austria},
    year = {2023},
    url = {https://www.R-project.org/},
  }

@article{rabiner1989,
  title={A tutorial on hidden {M}arkov models and selected applications in speech recognition},
  author={Rabiner, Lawrence R},
  journal={Proceedings of the IEEE},
  volume={77},
  number={2},
  pages={257--286},
  year={1989},
  publisher={Ieee},
  doi = {https://doi.org/10.1109/5.18626}
}

@article{rohani2003,
  title={Ecological interference between fatal diseases},
  author={Rohani, P and Green, CJ and Mantilla-Beniers, NB and Grenfell, BT},
  journal={Nature},
  volume={422},
  number={6934},
  pages={885--888},
  year={2003},
  publisher={Nature Publishing Group UK London},
  doi = {https://doi.org/10.1038/nature01542}
}

@article{rue2009,
author = {Rue, Håvard and Martino, Sara and Chopin, Nicolas},
title = {{Approximate Bayesian inference for latent Gaussian models by using integrated nested Laplace approximations}},
journal = {Journal of the Royal Statistical Society: Series B (Statistical Methodology)},
volume = {71},
number = {2},
pages = {319-392},
doi = {https://doi.org/10.1111/j.1467-9868.2008.00700.x},
year = {2009}
}

@article{rue2017,
  title={{Bayesian computing with INLA: A review}},
  author={Rue, H{\aa}vard and Riebler, Andrea and S{\o}rbye, Sigrunn H and Illian, Janine B and Simpson, Daniel P and Lindgren, Finn K},
  journal={Annual Review of Statistics and Its Application},
  volume={4},
  pages={395--421},
  year={2017},
  publisher={Annual Reviews},
  doi = {https://doi.org/10.1146/annurev-statistics-060116-054045}
}

@article{saarivirta2009,
    title={{Suomen terveydenhuoltojärjestelmän ja sairaaloiden kehittyminen. Vaatimattomista oloista modernin terveydenhuollon eturintamaan}},
    volume={3},
    url={https://journal.fi/kasvatusjaaika/article/view/68129},
    number={4},
    journal={Kasvatus \& Aika},
    author={Saarivirta, Toni and Consoli, Davide and Dhondt, Pieter},
    year={2009}
}

@article{saarivirta2012,
  title={{The evolution of the Finnish health-care system early 19th century and onwards}},
  author={Saarivirta, Toni and Consoli, Davide and Dhondt, Pieter},
  year={2012},
  journal={International Journal of Business and Social Sciences},
  volume={3},
  number={6},
  pages={243--257},
  publisher={Center for Promoting Ideas}
}

@article{speekenbrink2021,
      title={Ignorable and non-ignorable missing data in hidden {M}arkov models}, 
      author={Maarten Speekenbrink and Ingmar Visser},
      year={2021},
      journal={ArXiv preprint 2109.02770},
      doi={10.48550/arXiv.2109.02770},
      primaryClass={stat.ME}
}

@Misc{stan2024,
Author = {{Stan Development Team}},
Title = {Stan Modeling Language Users Guide and Reference Manual, version 2.34},
url ={https://mc-stan.org},
Year = {2024}
}

@article{tilastollinen1899,
    author = {{Tilastollinen päätoimisto}},
    year = {1899},
    title = {{Pääpiirteet Suomen väestötilastosta vuosina 1750-1890: 1: Väestön tila}},
    publisher = {Tilastollinen päätoimisto},
    journal = {SVT: Suomen virallinen tilasto VI. Väkiluvun-tilastoa 29.},
    address = {Helsinki, Finland},
    note = {Table 1 A}
}

@Article{vehtari2017,
    title = {{Practical Bayesian model evaluation using leave-one-out cross-validation and WAIC}},
    author = {Aki Vehtari and Andrew Gelman and Jonah Gabry},
    year = {2017},
    journal = {Statistics and Computing},
    volume = {27},
    issue = {5},
    pages = {1413--1432},
    doi = {https://doi.org/10.1007/s11222-016-9696-4},
  }

@article{vehtari2021,
author = {Aki Vehtari and Andrew Gelman and Daniel Simpson and Bob Carpenter and Paul-Christian B{\"u}rkner},
title = {Rank-Normalization, Folding, and Localization: An Improved $\widehat{R}$ for Assessing Convergence of {MCMC} (with Discussion)},
volume = {16},
journal = {Bayesian Analysis},
number = {2},
publisher = {International Society for Bayesian Analysis},
pages = {667--718},
year = {2021},
doi = {https://doi.org/10.1214/20-BA1221}
}

@article{vora2008,
author = {Vora, A and Burke, D S and Cummings, D A T},
title = {The impact of a physical geographic barrier on the dynamics of measles},
volume = {136},
DOI = {https://doi.org/10.1017/S0950268807009193},
number = {5},
journal = {Epidemiology and Infection},
year = {2008},
pages = {713–720}
}

@article{voutilainen2020,
  title={{A Bayesian reconstruction of a historical population in Finland, 1647--1850}},
  author={Voutilainen, Miikka and Helske, Jouni and H{\"o}gmander, Harri},
  journal={Demography},
  volume={57},
  number={3},
  pages={1171--1192},
  year={2020},
  publisher={Duke University Press},
  doi = {https://doi.org/10.1007/s13524-020-00889-1}
}

@article{vuorinen1999,
title = "Suomalainen tautinimist{\"o} ennen bakteriologista vallankumousta",
author = "Vuorinen, \{Heikki S\}",
year = "1999",
language = "suomi",
volume = "16",
pages = "33--61",
journal = "Hippokrates: Suomen l{\"a}{\"a}ketieteen historian seuran vuosikirja",
issn = "0781-5859",
publisher = "Suomen Laaketieteen Historian Seura",
}

@article{yeh2012,
author = {Hung-Wen Yeh and Wenyaw Chan and Elaine Symanski},
title = {Intermittent Missing Observations in Discrete-Time Hidden {M}arkov Models},
journal = {Communications in Statistics - Simulation and Computation},
volume = {41},
number = {2},
pages = {167--181},
year = {2012},
publisher = {Taylor \& Francis},
doi = {10.1080/03610918.2011.581778}
}

@article{yue2017,
  title={{Trade routes and plague transmission in pre-industrial Europe}},
  author={Yue, Ricci PH and Lee, Harry F and Wu, Connor YH},
  journal={Scientific reports},
  volume={7},
  number={1},
  pages={12973},
  year={2017},
  publisher={Nature Publishing Group UK London},
  doi = {https://doi.org/10.1038/s41598-017-13481-2}
}

@book{zucchini2009,
    author = {Zucchini, W. and MacDonald, I.L.},
    title = {Hidden Markov Models for Time Series: An Introduction Using R},
    publisher = {Chapman and Hall/CRC},
    year = {2009},
    doi = {https://doi.org/10.1201/9781420010893}
}

\end{document}